\documentclass[a4paper,11pt]{article}

\usepackage{jheppub} 

\usepackage[T1]{fontenc} 

\title{Active neutrino oscillations and double beta decay characteristics with
sterile neutrinos contributions}

\author
{V. V. Khruschov,}
\author
{S. V. Fomichev}
\author
{and S. V. Semenov}

\affiliation{National Research Centre Kurchatov Institute, Academician
Kurchatov Place 1, Moscow, 123182 Russia}

\emailAdd{Khruschov{\_}VV@nrcki.ru}
\emailAdd{Fomichev{\_}SV@nrcki.ru}
\emailAdd{Semenov{\_}SV@nrcki.ru}

\abstract{Light sterile neutrinos contributions both for active neutrinos
oscillations and neutrinoless double beta decay characteristics are estimated
in the framework of the phenomenological model with three active and three
sterile neutrinos assuming the Majorana nature of neutrino. Appearance and
survival probabilities for active neutrinos with contributions of eV-sterile
neutrinos in application to the short-baseline anomalies in neutrino data are
obtained at the same test values of the model parameters. Modified graphical
dependences for the survival probability of electron neutrinos/antineutrinos
and the probability of appearance of electron neutrinos/antineutrinos in muon
neutrino/antineutrino beams as functions of distance and other model
parameters at different neutrino energies, and also as functions of the
ratio of the distance to the neutrino energy are presented. A significant
difference was found between the probability curves of the considered neutrino
model and the simple sinusoidal curves of the neutrino model with one sterile
neutrino. Effective electron neutrino masses for the beta decay and the
neutrinoless double beta decay are estimated with account of eV-sterile
neutrinos contributions. Besides, the two-neutrino double beta decay
characteristics for selenium-82 are calculated. These findings can be used for
interpretation and prediction of results of ground-based experiments on search
for the sterile neutrinos as well as the neutrinoless double beta decay.}

\keywords{Neutrino Oscillation Anomalies at Short Distances, Sterile
Neutrinos, Double Beta Decay Characteristics}

\arxivnumber{}

\begin{document}
\maketitle
\flushbottom

\section{Introduction}
\label{Sec1}
At present there are indications to anomalies of neutrino fluxes for some
processes that cannot be explained using oscillation parameters only for
three active neutrinos in the framework of the Modified Standard Model (MSM).
MSM (or $\nu$SM) model is the Standard Model (SM) with three massive active
neutrinos instead of SM with massless neutrinos. These anomalies include the
LSND (or accelerator) anomaly (AA) \cite{Atha1996,Agu2001,Agu2013,Agu2018},
the reactor anomaly (RA or the reactor antineutrino anomaly RAA)
\cite{Mu2011,Me2011,Hu2011,ale18,ser18} and the gallium (calibration) anomaly
(GA) \cite{Abdu2009,Kae2010,Giunti2013}. These three types of short baseline
(SBL) anomalies  manifest themselves at short distances (more precisely, at
distances $L$ such that the numerical value of the parameter $\Delta m^2 L/E$,
where $E$ is the neutrino energy and $\Delta m^2$ is the characteristic
difference of the neutrino mass squares, is of the order of unity). For the
LSND anomaly \cite{Atha1996,Agu2001} an excess of the electron antineutrinos
in beams of muon antineutrinos in comparison with the expected value according
to the MSM is observed. Similar results were observed in the MiniBooNE
experiments for electron neutrinos and antineutrinos \cite{Agu2013,Agu2018}.
Deficit of reactor electron antineutrinos at short distances is called as RA,
while the deficit of electron neutrinos from a radioactive source occurred at
calibration of detectors for the SAGE and GALLEX experiments is commonly
called as GA. In other words, data on SBL anomalies refer to both the
inexplicable appearance of electron neutrinos or antineutrinos in beams of
muon neutrinos or antineutrinos, respectively, and to the disappearance of
electron neutrinos or antineutrinos at shot distances. SBL anomalies
may be explained by means of existence of one or two new sterile neutrinos
(SNs), which do not interact directly with the SM gauge bosons. The
characteristic mass scale of these light SNs is $1$~eV. Now intensive searches
are carried out for light SNs (or eV-sterile neutrinos). It is expected that
in the coming several years it will be possible to confirm or reject their
existence (see for example \cite{Abazajian2012,Gav2018,x35,x36,lyash}).

Adoption of SNs goes beyond MSM, so there have been proposed a number of
phe\-no\-me\-no\-lo\-gi\-cal models for prediction of effects due to them (see
for example \cite{Abazajian2012,KhruFom2016,Canetti2013,Conrad2013,%
KhruFom2017,PAZH2015,Warren,PAZH2016}). Models with additional SNs are usually
denoted as (3+$N$) models, or, in more detail, as ($k$+3+$n$+$m$) models,
where $k$ is the number of new neutrinos with masses less than masses of
active neutrinos, and $n$ and $m$ are the numbers of new neutrinos with masses
higher and considerably higher, respectively, than masses of the active
neutrinos \cite{Bilenky1977,Abazajian2012,Conrad2013,KhruFom2017,%
Schwetz2011,Kopp2013,Gariazzo2017,Bilenky}.

Now it is not yet known is the neutrino of the Dirac or Majorana fermion type
particle. Neutrino's type (or its nature) cannot be reveal in the neutrino
oscillation experiments \cite{PDG}. Practically the only process that allows
doing that is the neutrinoless double beta decay ($0\nu2\beta$-decay)
\cite{Radejo}. The study of double beta decay, both with the help of
theoretical investigations and creation of new large-scale installations is
now rapidly developing. The discovery of this rare process may indicate on
Majorana's nature of neutrino and will make it possible in this case to draw
conclusions about the neutrino mass matrix and mixing parameters.

Currently, the rarest phenomena that have been observed experimentally are
reactions of two-neutrino double beta decay ($2\nu2\beta$-decay) with
half-life time $T_{1/2}\sim 10^{19}-10^{24}$~years \cite{s1a}. The study of
the $2\nu2\beta$-decay allows ones to get information about the structure of
nuclei involved in the process, which is very essential for developing and
testing the nuclear models for theoretical consideration of neutrinoless
transitions. At the same time, two-neutrino channel is an unremovable
background for $0\nu2\beta$-decay and, therefore, the calculations of the
cor\-res\-pon\-ding differential intensities are needed for determining the
sensitivity of the experiments on delimitation of the effective Majorana
neutrino mass $m_{\beta\beta}$. Moreover, it is necessary also to take into
account the restriction on the effective neutrino mass $m_\beta$ that can be
measured in the tritium $\beta$-decay experiment KATRIN \cite{katrin}.

The mixing of neutrino states \cite{Bilenky1977} puts into operation with the
Pontecorvo--Maki--Nakagawa--Sakata matrix $U_{PMNS}\equiv U\equiv V\times P$,
so that $\psi_a^L=U_{ai}\psi_i^L$, where $\psi_{a,i}^L$ are left chiral fields
with flavor ``$a$'' or mass ``$m_i$'', $a=\{e,\mu,\tau\}$ and $i=\{1,2,3\}$. 
For three active neutrinos, the matrix $V$ is expressed in the standard
parametrization \cite{PDG} via the mixing angles $\theta_{ij}$ and the Dirac
CP phase, namely, the phase $\delta\equiv\delta_{\rm CP}$ associated with CP
violation in the lepton sector for Dirac or Majorana neutrinos, and
$P={\rm diag}\{1,e^{i\alpha},e^{i\beta}\}$, where
$\alpha\equiv\alpha_{\rm CP}$ and $\beta\equiv\beta_{\rm CP}$ are phases
associated with CP violation only for Majorana neutrinos. In oscillation
experiments it is impossible to measure $\alpha_{\rm CP}$ and
$\beta_{\rm CP}$. Nonetheless, in oscillation experiments the violation of
conservation laws for the lepton numbers $L_{e}$, $L_{\mu}$ and $L_{\tau}$ was
fixed.

Using the analysis of high-precision experimental data, the values of the
mixing angles and the differences of the neutrino mass squares
$\Delta m_{21}^2$ and $|\Delta m_{31}^2|$ (where
$\Delta m_{ij}^2=m_i^2-m_j^2$) were found \cite{PDG,Esteban2017}. CP phases
$\alpha_{\rm CP}$ and $\beta_{\rm CP}$ are currently unknown, and also unknown
is the order (hierarchy) of the neutrino mass spectrum, normal (NO) or inverse
(IO). Although the $\delta_{\rm CP}$ value is also not yet definitively
determined experimentally, in a number of papers its estimate was obtained
(see for example \cite{Esteban2017,KhruFom2016,Petkov}). For the NO-case
of the mass spectrum of active neutrinos (ANs), this estimate results in
$\sin\delta_{\rm CP}<0$ and $\delta_{\rm CP} \approx -\pi/2$. If we take into
account the restrictions on the sum of the neutrino masses from cosmological
observations \cite{Wang} and the results of the T2K experiment \cite{Kabe},
then the NO-case of the neutrino mass spectrum turns out to be preferable.
So in carrying out further numerical calculations, we restrict ourselves to
the NO-case, assuming $\delta_{\rm CP}=-\pi/2$.

The main purpose of this paper is to consider the effects of eV-sterile
neutrinos on oscillation properties of active neutrinos and neutrinoless
double-beta decay characteristics in the framework of the (0+3+3+0) neutrino
model, which we will refer as the (3+3) model, assuming neutrinos to be the
Majorana type particles. In section~\ref{Sec2} we briefly present the (3+3)
model with three ANs and three SNs for estimating SNs effects in the
$0\nu2\beta$-decay and ANs oscillation characteristics at small distances. The
propositions of this model have been outlined in the recent paper
\cite{KhruFom2019}.

However, in contrast to the sterile neutrino mass option considered in detail
in \cite{KhruFom2019}, where only one mass $m_4$ of sterile neutrino is of the
order of $1$~eV, while other two masses, $m_5$ and $m_6$, are essentially
heavier (the (3+2+1) model), in the present paper we consider neutrinos to be
Majorana particles, and therefore it is necessary to take into account
existing restrictions on the effective neutrino mass \cite{pet}, together with 
the values of mixing angles acceptable for describing the SBL anomalies. 
Therefore, in this paper, as an alternative case, the masses of all three SNs 
are selected in the range of about $1$~eV (the light SNs case or the (3+3) 
model). It results in the qualitative difference with respect to the results 
obtained in \cite{KhruFom2019}. In section~\ref{Sec3} the results of detailed 
calculations of the ANs oscillation characteristics at small distances, taking 
into account the light SNs effects, are presented. Calculations were carried 
out at selected test values of the model parameters, using the results 
obtained earlier \cite{KhruFom2016,KhruFom2017}. We hope that the results of 
these calculations can help to explain the experimental data on the SBL 
neutrino anomalies.

In section~\ref{Sec4}, the estimations of the effective neutrino masses for
the beta decay and the neutrinoless double beta decay with account of SNs
contributions are performed using the (3+3) model parameter values. In
appendix~\ref{SecA}, the calculation of an amplitude for the
$2\nu2\beta$-decay within the High-States Dominance (HSD) and Single-State
Dominance (SSD) mechanisms for $^{82}$Se are presented. In section~\ref{Sec5}
the obtained results are accentuated.

\section{Some provisions of a (3+3) model for active and sterile neutrinos}
\label{Sec2}
Below, a (3+3) model is used to study the effects of SNs. This model was
detailed in \cite{KhruFom2019} and includes three AN $\nu_a\,(a=e,\mu,\tau)$
and three new neutrinos: a sterile neutrino $\nu_s$, a hidden neutrino $\nu_h$
and a dark neutrino $\nu_d$. A $6\times 6$ mixing matrix $U_{\rm mix}$ is used
in the model framework. For the compactness of the formulas, the symbols $h_s$
and $h_{i'}$ are introduced for additional left flavor and mass fields,
respectively. Above, $s$ denotes a set of indices that highlight the fields
$\nu_s$, $\nu_h$ and $\nu_d$ among $h_s$, while $i'$ denotes a set of indices
$4$, $5$ and $6$. Then matrix $U_{\rm mix}$, which establishes linkage between
the flavor and mass neutrino fields, is represented as:
\begin{equation}
\left(\begin{array}{c}
\nu_a\\ h_s \end{array}\right)=U_{\rm mix}\left(\begin{array}{c}
\nu_i\\ h_{i'}\end{array}\right)\equiv
\left(\begin{array}{cc}
\varkappa U&\sqrt{1-\varkappa^2}\,a\\
\sqrt{1-\varkappa^2}\,bU&\varkappa c\end{array}\right)
\left(\begin{array}{c}
\nu_i\\ h_{i'}\end{array}\right).
\label{eq21}
\end{equation}
Here $\varkappa=1-\epsilon$, where $\epsilon$ is a small quantity,
$U\equiv U_{\rm PMNS}$, where $U_{\rm PMNS}$ is the well-known unitary
$3\times 3$ mixing matrix of ANs ($U_{\rm PMNS}U_{\rm PMNS}^+=I$). Moreover,
$a$ and $b$ are arbitrary unitary $3\times 3$ matrices, with $c=-b\times a$.
Matrix $U_{\rm mix}$ under these conditions is unitary. We will use the
following matrices $a$ and $b$, which first were proposed in
\cite{KhruFom2016}:
\begin{equation}
a=\left(\begin{array}{lcr}
\,\,\,\,\,\cos\eta_2 & \sin\eta_2 & 0\\
-\sin\eta_2 & \cos\eta_2 & 0\\
\qquad 0 & 0 & e^{-i\kappa_2}\end{array}\right),\quad
b=-\left(\begin{array}{lcr}
\,\,\,\,\,\cos\eta_1 & \sin\eta_1 & 0\\
-\sin\eta_1 & \cos\eta_1 & 0\\
\qquad 0 & 0 & e^{-i\kappa_1}\end{array}\right).
\label{eq22}
\end{equation}
The values of mixing angles $\theta_{ij}$ of ANs for $U_{\rm PMNS}$ matrix
will be taken from the conditions $\sin^2\theta_{12} \approx 0.297$,
$\sin^2\theta_{23} \approx 0.425$ and $\sin^2\theta_{13} \approx 0.0215$
\cite{PDG}. For additional mixing parame\-ters concerning SNs, the following
trial values have been proposed: $\kappa_1=\kappa_2=-\pi/2$ and $\eta_1=5^o$,
while the matched values for $\eta_2$ and $\epsilon<0.03$ will be concretized
in the next Section.

Neutrino masses are given by a set of values $\{m\}=\{m_i,m_{i'}\}$ in eV
units: $m_1\approx 0.0016$, $m_2\approx 0.0088$ and $m_3\approx 0.0497$
\cite{KhruFom2016,PAZH2016}. In \cite{KhruFom2019}, it was used the value
$m_4\approx 1$~eV and the values $m_5$ and $m_6$ of the order of several keV.
Such values lead to rapidly oscillating curves for the probabilities of
conservation and appearance of electron neutrinos/antineutrinos that have to
be averaged to make comparison with the experimental data. This situation
persists until the masses $m_5$ and $m_6$ are of the order of $10$~eV. For
$m_5$ and $m_6$ values noticeably less than $10$~eV, the curves become
smoothly oscillating. In the present work, we use the values $m_4=1.05$~eV,
$m_5=0.63$~eV and $m_6=0.27$~eV. They are consistent with the acceptable mass
range found by treatment of the data relevant to SBL anomalies \cite{x34}.
However, the values of $m_5$ and $m_6$ are much less than the corresponding
values used in \cite{KhruFom2019}. This is due to the fact that the case of
Majorana neutrinos will be inspected below, for which $0\nu2\beta$-decay is
possible. So, we should take into account existing restrictions on the
effective Majorana neutrino mass $m_{\beta\beta}$ along with acceptable values
of angles $\theta_{\mu e}$ and $\theta_{ee}$ taken from global processing of
experimental data on SBL anomalies \cite{x35,x36,x34}, as well as restrictions
on the effective $\beta$-decay neutrino mass $m_{\beta}$ \cite{katrin}.

The probability amplitudes for describing the oscillations of neutrino flavors
can be found by solving well-known equations for the neutrino propagation (see
for example \cite{KhruFom2016,BS2013}). Using these equations, analytical
expressions for probabilities of different transitions between neutrino
flavors in beams of neutrinos/antineutrinos with energy $E$ in vacuum can be
obtained as functions of the distance $L$ from the neutrino source
\cite{Bilenky}. If $\widetilde{U}\equiv U_{\rm mix}$ is a generalized
$6\times 6$ matrix of neutrino mixing in the form of equation (\ref{eq21}),
then, using the notation $\Delta_{ki}\equiv \Delta m_{ik}^2 L/(4E)$, we can
calculate the probabilities of transitions from
$\nu_{\alpha}$ to $\nu_{\alpha'}$ or from $\bar{\nu}_\alpha$ to
$\bar{\nu}_{\alpha'}$ according to the formula
\begin{eqnarray}
P(\nu_{\alpha}(\overline{\nu}_{\alpha})\rightarrow\nu_{\alpha^{\prime}}
(\overline{\nu}_{\alpha^{\prime}}))=\delta_{\alpha^{\prime}\alpha}
&-4\sum_{i>k}{\rm Re}(\widetilde{U}_{\alpha^{\prime} i}
\widetilde{U}_{\alpha i}^{\ast}\widetilde{U}_{\alpha^{\prime} k}^{\ast}
\widetilde{U}_{\alpha k})\sin^2\Delta_{ki}\,\nonumber \\
&\pm2\sum_{i>k}{\rm Im}(\widetilde{U}_{\alpha^{\prime} i}
\widetilde{U}_{\alpha i}^{\ast}\widetilde{U}_{\alpha^{\prime} k}^{\ast}
\widetilde{U}_{\alpha k})\sin 2\Delta_{ki}\,,
\label{eq23}
\end{eqnarray}
where the upper sign (+) corresponds to neutrino transitions
$\nu_\alpha \to \nu_{\alpha'}$ and the lower sign (--) corresponds to
antineutrino transitions $\bar{\nu}_\alpha \to \bar{\nu}_{\alpha'}$. Note that
flavor indexes $\alpha$ and $\alpha'$ (as well as summation indices over
massive states $i$ and $k$) apply to both ANs and SNs.

\section{Numerical results for oscillation characteristics of active neutrinos
taking into account their mixing with eV-sterile neutrinos}
\label{Sec3}

\begin{figure}[htbp]
\center
\includegraphics[width=0.8\textwidth]{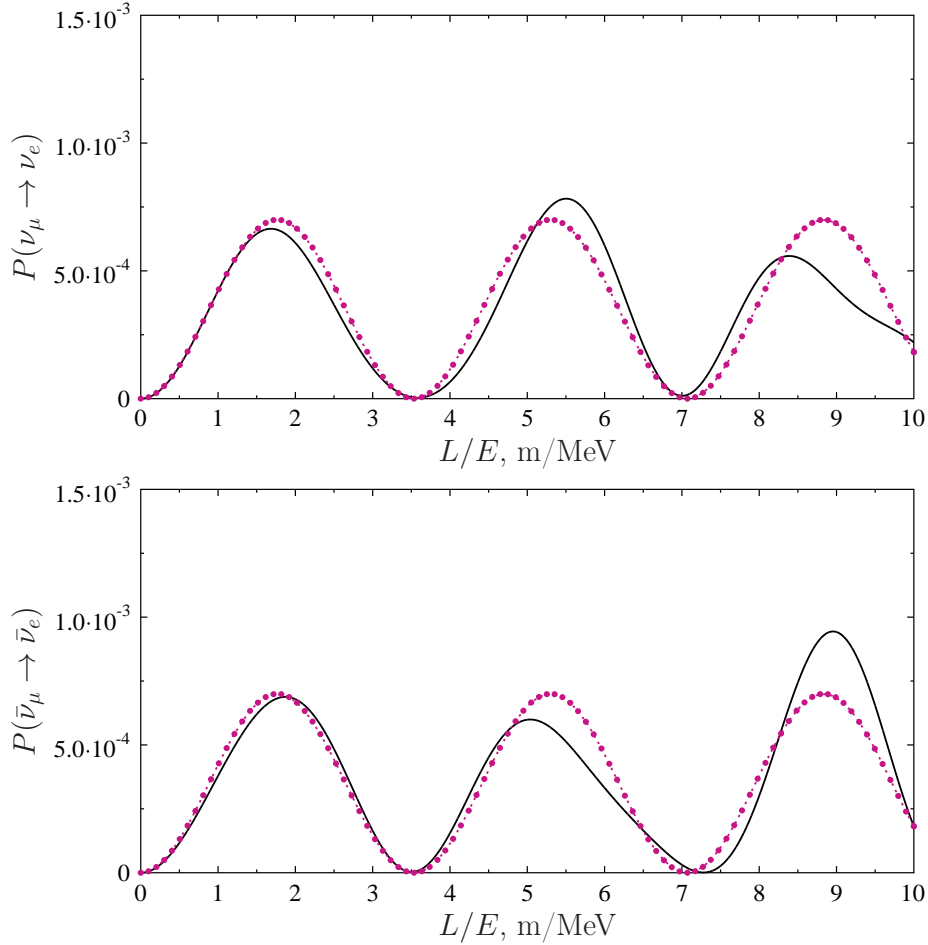}
\caption{The probability of appearance of $\nu_e$ (top panel) and
$\bar{\nu}_e$ (bottom panel) depending on the ratio of the distance $L$ from
the source to the neutrino energy $E$ in the beams of $\nu_\mu$ and
$\bar{\nu}_\mu$, respectively. For matrix $U_{\rm mix}$, $\epsilon=0.015$ and
$\eta_2=\pi/3$. The mass in square differences are
$\Delta m_{41}^2=1.1$~eV$^2$ and $\Delta m_{51}^2=0.4$~eV$^2$. The dotted
curves with circle marks that are the same in both panels and closely
approximate the solid curves show the probability values calculated in the
approximation of two neutrino states in the (3+1) model for
$\sin^2(2\theta_{\mu e})=0.0007$ and $\Delta m_{41}^2=0.7$~eV$^2$.}
\label{fig1}
\end{figure}

\begin{figure}[htbp]
\center
\includegraphics[width=0.7\textwidth]{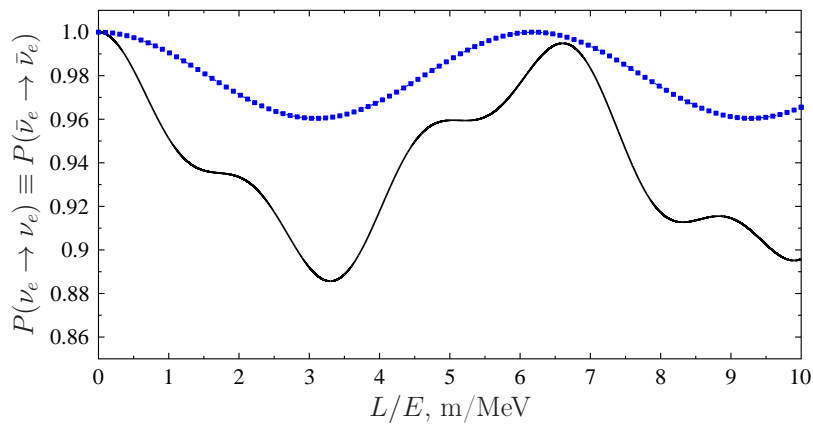}
\caption{The survival probabilities for $\nu_e$ ($\bar{\nu}_e$) depending on
the ratio of the distance $L$ from the source to the neutrino energy $E$ in
the beams of $\nu_{e}$ ($\bar{\nu}_e$). For the matrix $U_{\rm mix}$,
$\epsilon=0.015$ and $\eta_2=\pi/3$. The mass in square differences are
$\Delta m_{41}^2=1.1$~eV$^2$ and $\Delta m_{51}^2=0.4$~eV$^2$. The dotted
curve with square marks shows the probability values calculated in the
approximation of two neutrino states in the (3+1) model for the parameter
values $\sin^2(2\theta_{ee})=0.0396$ and $\Delta m_{41}^2=0.4$~eV$^2$, which
were obtained by joint processing the experimental data on RAA and GA.}
\label{fig2}
\end{figure}

\begin{figure}[htbp]
\center
\includegraphics[width=0.95\textwidth]{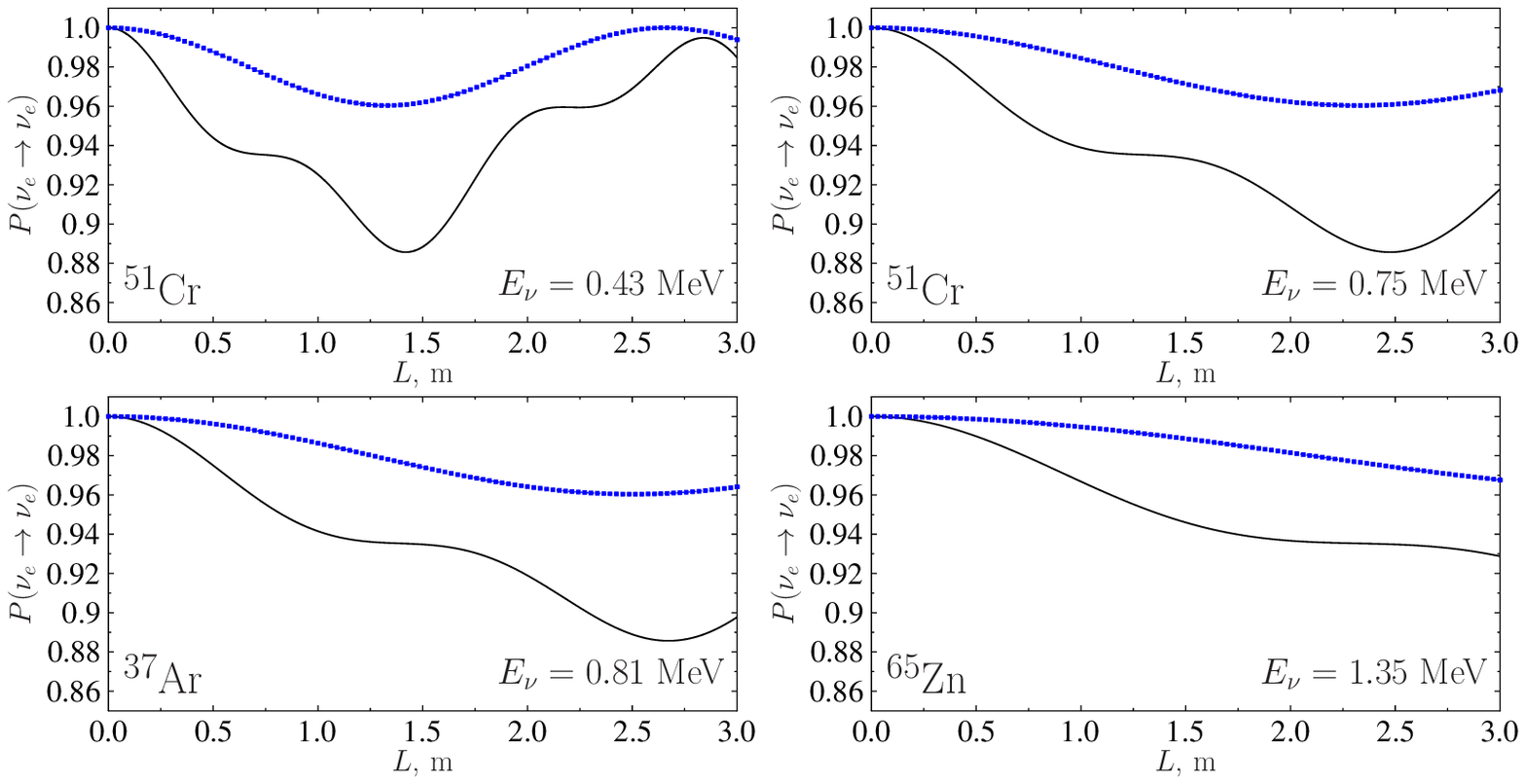}
\caption{The $\nu_e$ survival probability depending on the distance $L$ from
the source, in the capacity of which are the nuclei of various elements with
different neutrino energy, for the case of the mixing matrix $U_{\rm mix}$
with $\epsilon=0.015$ and $\eta_2=\pi/3$. The differences of the squared
masses are $\Delta m_{41}^2=1.1$~eV$^2$ and $\Delta m_{51}^2=0.4$~eV$^2$. The
dotted curves with square marks show the probability values calculated in the
approximation of two neutrino states of a (3+1) model for parameter values
$\sin^2(2\theta_{ee})=0.0396$ and $\Delta m_{41}^2=0.4$~eV$^2$ taken from
global processing of experimental data on GA and RAA \cite{x35,x36,x34}.}
\label{fig3}
\end{figure}

In this section a comparison is made between survival and appearance
neutrino/antineutrino probabilities calculated in the framework of the (3+3)
model and similar quantities calcula\-ted in the framework of the (3+1) model
based on currently available data for SBL anomalies \cite{x35,x36,x34}
(see figures~\ref{fig1}, \ref{fig2} and \ref{fig3}). First of all it refers to
data on the disappearance of muon neutrinos and antineutrinos and the
appearance of electron neutrinos and antineutrinos in the processes
$\overline{\nu}_\mu\to\overline{\nu}_e$ and $\nu_\mu\to\nu_e$. The typical
ratio of the distance traversed by the neutrino before detection to the
neutrino energy is either a few meters per MeV, or one meter per several MeVs.
Attempts of the simultaneous description of all the data in these processes
lead to difficulties. In particular, the problems associated with different
values of the excess of the output $\nu_e$ and $\overline{\nu}_e$ in the
MiniBooNE experiment can be reduced under the condition of CP violation
\cite{Mal2007,Pal2005,Kar2007}.

It is also possible to describe the reactor and gallium anomalies  in the
framework of the model considered  by selecting the appropriate value of the
parameter $\epsilon$ (see equation \ref{eq21}). It is sufficient to choose the
value corresponding to the experimental data for the parameter
$\varkappa=1-\epsilon$, that naturally leads to the deficit of electron
neutrinos and antineutrinos. Note that the status of RA with allowance for the
recently discovered excess of the number of antineutrinos in comparison with
the model calculations in the $5$~MeV range and thus confirmation of the
possible existence of a light SN with a mass of about $1$~eV is discussed for
example in Refs. \cite{Dentler,Dan,Gari}.

Figure~\ref{fig1} shows the appearance probabilities of $\nu_e$ (upper panel)
and $\bar{\nu_e}$ (lower panel) in the beams of $\nu_{\mu}$ and
$\bar{\nu}_{\mu}$, respectively, as a function of the ratio of distance from
the source to the neutrino energy at the coupling constant $\epsilon=0.015$
and the parameter $\eta_2=\pi/3$ and for the considered in this paper neutrino
mass spectrum. In this case, the contribution of sterile neutrinos has the
character of smooth oscillations. Furthermore, for example, already with the
value of $\epsilon=0.005$ the relative outputs of $\nu_e$ and
$\overline{\nu}_e$ increase by approximately two orders of magnitude
(up to $\sim 10^{-4}$) in comparison with their values for $\epsilon=0$
($\sim 10^{-6}$). The comparison with the experiment is provided with the help
of simple (3+1) model using the formula $P_{3+1}(\nu_{\mu}\to\nu_e)
=P_{3+1}(\bar{\nu}_{\mu}\to\bar{\nu}_e)
=\sin^2(2\theta_{\mu e})\sin^2(1.27\Delta m_{41}^2L/E)$, where $L$ is the
distance to the detector in meters, $E$ is the neutrino energy in MeV and
$\Delta m_{41}^2$ is the difference between the squared neutrino masses in
eV$^2$.


In figure~\ref{fig2}, the similar results are given for
$P(\nu_e\to\nu_e)\equiv P(\overline{\nu}_e\to\overline{\nu}_e)$ for the same
values of the model parameters. The new salient feature of our model curve
(solid line) consists in the strong difference from the sinusoidally varying
curve, which can be only obtained by the standard formula within the (3+1)
model for probability $P(\nu_e\to\nu_e)$, that is, by the formula
$P_{3+1}(\nu_e\to\nu_e)=1-\sin^2(2\theta_{ee})\sin^2(1.27\Delta m_{41}^2L/E)$.
Moreover the periods of variation for the model curves presented in
figures~\ref{fig1} and \ref{fig2} are different in spite of the same neutrino
mass spectrum. So it is permissible to argue for effectively varying SN mass
values in different processes. This fact can shed light on the possible
obstacle in determining a precise value of a SN mass.

As for the possible description of the gallium anomaly within the (3+3) model
framework, the survival probability for $\nu_e$ as a function of distance $L$
from the neutrino source for various values of the neutrino energy $E$ is
shown in figure~\ref{fig3} for the same parameters $\epsilon=0.015$ and
$\eta_2=\pi/3$ of the mixing matrix and for the same mass values as it was
used in figures~\ref{fig1} and \ref{fig2}. In each panel of figure~\ref{fig3},
that value of neutrino energy $E$ is selected, which is acquired by electron
neutrino in the process involving artificial sources containing isotopes
$^{51}$Cr, $^{37}$Ar and $^{65}$Zn, respectively. SNs contributions have
character of smooth oscillation. Moreover, these oscillations coincide in
phase with the oscillations, which are obtained by the standard formula of the
(3+1) model for probability $P(\nu_e\to\nu_e)$.

The results obtained are characteristic features of the (3+3) neutrino model
with SNs contributions, which is considered in the present paper, and they can
be used for interpreting both the available and coming experimental data on
the SNs search.

\section{Contributions of eV-sterile neutrinos to the probability of beta
decay and neutrinoless double beta decay}
\label{Sec4}
If the resolution of the experimental setup does not allow to distinguish
between the masses of massive neutrinos involved in beta decay, then in this
case the weighted mass of the electron neutrino is used, which is called as
effective neutrino mass $m_{\beta}$ \cite{katrin}:
\begin{equation}
m_{\beta}^2=\sum\nolimits_i|U_{ei}|^2 m_i^2.
\label{eq41}
\end{equation}
We will use as $\{m_i\}$ the set of masses suggested in the previous section
along with the matrix elements $U_{ei}$ used there. The value of $m_{\beta}$
calculated in this way is $0.131$~eV. Note that for now this value cannot be
detected in the current KATRIN experiment, for which the achievable lower
bound of the effective mass of electron neutrino is estimated as $0.2$~eV (at
$90\%$ CL) \cite{kat}. Furthermore, the value of $m_{\beta}$ is about
$0.01$~eV with only three ANs and without SNs contribution.

To calculate the probability of neutrinoless double beta decay due to exchange
by light massive Majorana neutrinos, it is necessary to take into account
their contributions with allowance for the Majorana phases \cite{gerda}:
\begin{equation}
m_{\beta\beta}=\left|\sum\nolimits_iU_{ei}^2 m_i\right|.
\label{eq42}
\end{equation}
Since the values of the Majorana phases are still not known, we estimate the
maximum value of the effective Majorana mass of an electron neutrino using
the set $\{m_i\}$ proposed in the previous section along with the same matrix
elements $U_{ei}$ from there. Then the value of $m_{\beta\beta}$ is equal to
$0.027$~eV. This value is approximately half of the currently most 
constraining upper limit on effective Majorana mass of electron neutrino, 
which is obtained in the KamLAND-Zen experiment with $^{136}$Xe ($0.061$~eV) 
\cite{pet}. Notice that $m_{\beta\beta}$ is equal to $0.0026$~eV when only 
contribution of ANs is taken into account.

For the half-life with respect to the neutrinoless decay, the expression
takes place as
\begin{equation}
T_{1/2}^{0\nu}=\frac{m_e^2}
{G^{0\nu}g_A^4\left|M^{0\nu}m_{\beta\beta}\right|^2}.
\label{eq43}
\end{equation}
Here $G^{0\nu}$ is the phase-space factor, $M^{0\nu}$ is the nuclear 
transition matrix element, and $m_e$ is the electron mass.

Now there is an intensive searches for $0\nu2\beta$-decay and plans are
underway for a number of large-scale experiments such as SuperNEMO, LEGEND,
EXO, CUPID-0, etc. $^{82}$Se is the one of perspective stable isotopes for
these investigations, which is used in recent CUPID-0 and SuperNEMO projects.
So, it is of great interest to perform calculations of double beta-decay
characteristics of $^{82}$Se. Namely for this isotope, $2\nu2\beta$-decay for
the first time was observed in a direct experiment in 1987. In NEMO-3
experiment, where $0.93$~kg of $^{82}$Se were used as the $2\beta$-source
\cite{s3}, the total and differential intensities for the two-neutrino channel
were measured and low bound for half-decay time of neutrinoless transition was
determined: $T_{1/2}^{0\nu}>2.5\!\times\!10^{23}$~years. In CUPID-0 setup
effectively $5.53$~kg of $^{82}$Se are involved in data recording, and the
following result was obtained \cite{s2a,s2b,s2c}:
$T_{1/2}^{0\nu}>2.4\!\times\!10^{24}$~years. Phase factor calculations
\cite{s4} and existing models of nuclear structure \cite{s5} lead to the
following conclusions for $^{82}$Se:
$G^{0\nu}=10.16\!\times\!10^{-15}$~year$^{-1}$,
$\left|M^{0\nu}\right|_{\rm min}=2.64$,
$\left|M^{0\nu}\right|_{\rm max}=4.64$.
Taking into account the above estimate of $m_{\beta\beta}$, we can calculate
the values of $T_{1/2}^{0\nu}$ for the stable isotope $^{82}$Se:
$\{T_{1/2}^{0\nu}\}_{\rm min}=6.17\!\times\!10^{26}$~years,
$\{T_{1/2}^{0\nu}\}_{\rm max}=1.96\!\times\!10^{27}$~years.
Thus, it is necessary to increase significantly the sensitivity of the
experiment on search for neutrinoless double beta decay of $^{82}$Se. The new
large-scale installation SuperNEMO \cite{s6}, where it is planned to use up to
$100$~kg of $^{82}$Se, serves this purpose. As noted above, for evaluation of
sensitivity of the planned experiment, it is necessary to determine precisely
the energy spectrum of electrons arising in a two-neutrino channel that will
be done in appendix~\ref{SecA}.

\section{Discussion and conclusions}
\label{Sec5}
In this paper, the light SNs contributions both to ANs oscillations at short
distances and neutrinoless double beta decay characteristics are estimated in
the framework of the phe\-no\-me\-no\-lo\-gi\-cal (3+3) neutrino model. The
properties of these characteristics at the test values of the model
parameters are numerically investigated. All calculations were performed for
the case of a normal hierarchy of the mass spectrum of active neutrinos with
allowance for possible violation of the CP invariance in the lepton sector and
for the value $-\pi/2$ for the Dirac CP phase in the $U_{\rm PMNS}$ matrix. In
section~\ref{Sec3}, graphical dependences of the probabilities of appearance
and disappearance of electron neutrinos and antineutrinos pertaining to both
acceleration and reactor/gallium anomalies, respectively, are exhibited as a
functions of the ratio of the distance from the neutrino source to the
neutrino energy. Significant difference of the probability curves of our (3+3)
neutrino model from simple sinusoidal curves of the (3+1) neutrino model,
which is clearly seen in figures~\ref{fig1}, \ref{fig2} and \ref{fig3} is one
of the main results of the present work. Such behavior of the survival and 
appearance probabilities of electron neutrinos and antineutrinos may be a 
cause of difficulties concerning with a valid confirmation of neutrino 
anomalies at short distances. Besides, the considered (3+3) neutrino model 
gives possibility to describe the data of all SBL neutrino anomalies with the 
same model parameters, that is with the same mixing parameters and the mass 
values pertaining to the eV-sterile neutrinos, that cannot be reached in the 
scope of the simple (3+1) neutrino model. So, a global self-consistent 
description of the neutrino data for all SBL anomalies on the basis of the 
simple (3+1) model of neutrino may not be possible.

The results obtained make it possible to interpret the experimental data on
oscillations of neutrinos and antineutrinos that admit the existence of LNSD
and MiniBooNE anomalies, as well as the reactor antineutrino anomaly together
with the gallium anomaly. The most promising test for searching light SNs is,
as noted above, a verification of the existence of a gallium anomaly. In any
case the experimental data on SBL anomalies should be processed separately for
the every anomaly and also separately for neutrino and antineutrino data for
acceleration anomaly.

In section~\ref{Sec4}, estimates of the effective masses $m_{\beta}$ and
$m_{\beta\beta}$ of the electron neutrino taking into account the sterile
neutrino contributions are obtained, which can be tested in experiments on
$\beta$-decay and $0\nu2\beta$-decay, in particular, in the KATRIN, SuperNEMO
and CUPID experiments. Besides, an analysis of the energy structure of
$2\nu2\beta$-decay of $^{82}$Se made in appendix~\ref{SecA} results in the
conclusion that this process can be considered as a virtual two-step
transition, which connects the initial and final states through the first
$1^+$-state of the intermediate nucleus (that is the Single State Dominance
mechanism -- SSD). For $^{82}$Se, this state is excited $1_1^+$-state of
$^{82}$Br with $E_x=75$~keV, since the quantum number of the ground state of
bromine-82 is $J^\pi=5^-$. In this paper, we obtain estimates of the half-life
times for neutrinoless and two-neutrino double beta decays of $^{82}$Se,
respectively, in section~\ref{Sec4} and appendix~\ref{SecA}. Measurement of
the half-life time $T_{1/2}^{2\nu2\beta}$ enables to calculate theoretically
the value of a nuclear matrix element
$M_1^F=\langle 0_f^+\|\hat{\beta}^-\|1_1^+\rangle$, which has not yet been
determined experimentally from the recharge reaction. In addition, the
calculated distribution in energy of one electron will give possibility to
establish, how it is planned during the project SuperNEMO, what mechanism, SSD
or HSD, is valid for a two-neutrino channel (see figure~\ref{fig4}). This is
important both for calculating an unrecoverable background and for evaluating
the sensitivity of the experiment on search of the neutrinoless double beta
decay of isotope $^{82}$Se.

\appendix
\section{The amplitude of two-neutrino double beta decay of selenium-82}
\label{SecA}
Let us carry out the theoretical calculations of the total and differential
intensities of $2\nu2\beta$-decay of $^{82}$Se. To calculate the intensity of
two-neutrino transitions one needs to perform summation over all possible
$1^{+}$-states of the intermediate nucleus \cite{s7,s8}. For this, it is
necessary to know the values of modules and phases of matrix elements
$\langle 0_f^+\|\hat{\beta}^-\|1_N^+\rangle$ and
$\langle 1_N^+\|\hat{\beta}^-\|0_i^+\rangle$,
with $\hat{\beta}^-=\sigma\tau^-$, which are involved in the expression for
$T_{1/2}^{2\nu2\beta}$:
\begin{align}
\left[T_{1/2}^{2\nu 2\beta}\left(0_i^+ \to 0_f^+\right)\right]^{-1}
&=\frac{G_\beta^4g_A^4}{32\pi^7\ln 2}\int\limits_{m_e}^{T+m_e}{d\varepsilon_1}
\int\limits_{m_e}^{T+2m_e-\varepsilon_1}d\varepsilon_2
\int\limits_0^{T+2m_e-\varepsilon_1-\varepsilon_2}d\omega_1\times\nonumber \\
&\times F(Z_f,\varepsilon_1)F(Z_f,\varepsilon_2)p_1\varepsilon_1 p_2
\varepsilon_2\omega_1^2\omega_2^2 A_{0_f^+}.
\label{eqA1}
\end{align}
The expression for $A_{0_f^+}$ is as follows:
\begin{align}
4A_{0_f^+}&=\bigg|\sum\limits_N\left\langle 0_f^+\left\|\hat{\beta}^-\right\|
1_N^+\right\rangle\left\langle 1_N^+\left\|\hat{\beta}^-\right\|0_i^+
\right\rangle(K_N+L_N)\bigg|^2+ \nonumber \\
&+\frac{1}{3}\bigg|\sum\limits_N\left\langle 0_f^+\left\|\hat{\beta}^-\right\|
1_N^+\right\rangle\left\langle 1_N^+\left\|\hat{\beta}^-\right\|0_i^+
\right\rangle(K_N-L_N)\bigg|^2.
\label{eqA2}
\end{align}
Here $p_1$, $p_2$ and $\varepsilon_1$, $\varepsilon_2$ are, respectively,
momentums and energies of the electrons, $\omega_1$ and
$\omega_2=T+2m_e-\varepsilon_1-\varepsilon_2-\omega_1$ are
antineutrino energies, $T=E_i-E_f-2m_e=Q_{\beta\beta}$ is the total kinetic
energy of leptons in final state, and $E_i(E_f)$ is the mass of the parent
(daughter) nucleus. Then, $F(Z_f,\varepsilon)$ is Coulomb factor taking into
account the influence of electrostatic field of the nucleus to the emitted
electrons, while the quantities $K_N$ and $L_N$ contain the energy
denominators of the second-order perturbation theory:
\begin{subequations}
\begin{align}
&K_N=\frac{1}{\mu_N+\left(\varepsilon_1+\omega_1-\varepsilon_2-\omega_2\right)
/2}+\frac{1}{\mu_N-\left(\varepsilon_1+\omega_1-\varepsilon_2-\omega_2\right)
/2}, \\
&L_N=\frac{1}{\mu_N+\left(\varepsilon_1+\omega_2-\varepsilon_2-\omega_1\right)
/2}+\frac{1}{\mu_N-\left(\varepsilon_1+\omega_2-\varepsilon_2-\omega_1\right)
/2},
\label{eqA3}
\end{align}
\end{subequations}
with $\mu_N=E_N-(E_i+E_f)/2$ and $E_N$ the energy of the $N$-th $1^+$-state
of intermediate nucleus.

Calculation of nuclear matrix elements
$\langle 0_f^+\|\hat{\beta}^-\|1_N^+\rangle$ and
$\langle 1_N^+\|\hat{\beta}^-\|0_i^+\rangle$ is a very difficult theoretical
problem \cite{s9}. At the same time, it can be assumed that for some isotopes
the main contribution in the sums over $N$ in the expression (\ref{eqA2}) is
the contribution of the ground state of the intermediate nucleus in the case
if this state has the quantum number $J^\pi=1^+$. This two-neutrino double
beta decay mechanism meets the hypothesis of dominance of the ground state of
the intermediate nucleus (SSD mechanism -- Single State Dominance
\cite{s10,s11}). This situation occurs for $^{100}$Mo, where the
$2\nu2\beta$-transition can be considered with a good accuracy as a two-step
process, which links the initial ($^{100}$Mo) and final ($^{100}$Ru) states of
the process via the ground $1^+$-state of the intermediate nucleus $^{100}$Tc.
Nuclear matrix elements
$M_1^I=\langle 1_{g.s.}^+\|\hat{\beta}^-\|0_i^+\rangle$ and
$M_1^F=\langle 0_f^+\|\hat{\beta}^-\|1_{g.s.}^+\rangle$
can be found from the force transition values $ft$ for the electron capture
or single beta decay process. Here $ft$ is the product of the phase factor
by the half-life time of the corresponding single $\beta$-process. So,
\begin{equation}
M_1^I=\frac{1}{g_A}\sqrt{\frac{3D}{ft_{EC}}}, \quad
M_1^F=\frac{1}{g_A}\sqrt{\frac{3D}{ft_{\beta^-}}},
\label{eqA4}
\end{equation}
where $g_A=1.27561$, $D=\frac{2\pi^3\ln 2}{G_\beta^2 m_e^5}=6288.6$~s,
$G_\beta=G_F\cos\theta_C$, $G_F=1.166378\cdot10^{-5}$~GeV$^{-2}$, and
$\cos\theta_C=0.97425$.

If the SSD hypothesis is valid under the condition that the quantum number of
the ground state of the intermediate nucleus is $J^\pi=1^+$, in this case the
intensity of the two-neutrino transition is determined only by the
intensities of single beta processes, which are characterized by factors
$ft_{\beta^-}$ and $ft_{EC}$, and is independent of $G_\beta$ and $g_A$
\cite{s12}:
\begin{equation}
T_{1/2}^{(2\nu)}(0^+ \to 0^+)=\frac{16\pi^2 ft_{EC}ft_{\beta^-}}
{3\ln 2\,(\lambda_C/c)H(T,0_f^+)}=2.997\times 10^{14}years\times
\frac{10^{\log ft_{EC}+\log ft_{\beta^-}}}{H(T,0_f^+)},
\label{eqA5}
\end{equation}
where
\begin{align}
H(T,0_f^+)&=\int\limits_1^{T+1}d\varepsilon_1\int\limits_1^{T+2-\varepsilon_1}
d\varepsilon_2\int\limits_0^{T+1-\varepsilon_1-\varepsilon_2}d\omega_1 \times
\nonumber \\
&\times F(Z_f,\varepsilon_1)F(Z_f,\varepsilon_2)p_1\varepsilon_1 p_2
\varepsilon_2\omega_1^2\omega_2^2(K^2+KL+L^2).
\label{A6}
\end{align}
The value of $\log ft_{\beta^-}$ is well established from the beta decay of
$^{100}$Tc and is equal to $4.59$ that corresponds to $M_1^F=0.546$.
Definition of $\log ft_{EC}$ from experiments on the study of electron capture
in $^{100}$Tc is a difficult experimental task. Currently the most
accurate value of $\log ft_{EC}$ for electron capture in $^{100}$Tc was
obtained in \cite{s13}, namely, $\log ft_{EC}=4.29_{-0.07}^{+0.08}$.

When calculating the half-life time for a $2\nu2\beta$ transition, often it is
assumed that the kinetic energies of the emitted leptons are approximately
equal \cite{s7,s8,s14}. Then $K\approx L\approx 2/\mu$. This situation is
equivalent by dominance in the expression for $T_{1/2}^{2\nu2\beta}$ of
contributions related to the intermediate nucleus states with high excitation
energy (HSD mechanism -- High States Dominance). As has been shown in
\cite{s15}, such an approach, when the dependence of $K$ and $L$ on the lepton
energies is neglected, leads to overestimating the theoretical value of
$T_{1/2}^{2\nu2\beta}$. In the case of $0^+\to 0_{g.s.}^+$ transition in
$^{100}$Mo, the effect is about 25\%. Taking into account the dependence of
$K$ and $L$ on lepton energies on the basis of the SSD mechanism allows one to
obtain differential intensity in energy of one electron
$P(\varepsilon)=d\ln I/d\varepsilon$ for the $2\nu2\beta$-decay of the isotope
$^{100}$Mo \cite{s16a,s16b}, which matches the NEMO-3 data \cite{s17}.

\begin{figure}[htbp]
\center
\includegraphics[width=0.7\textwidth]{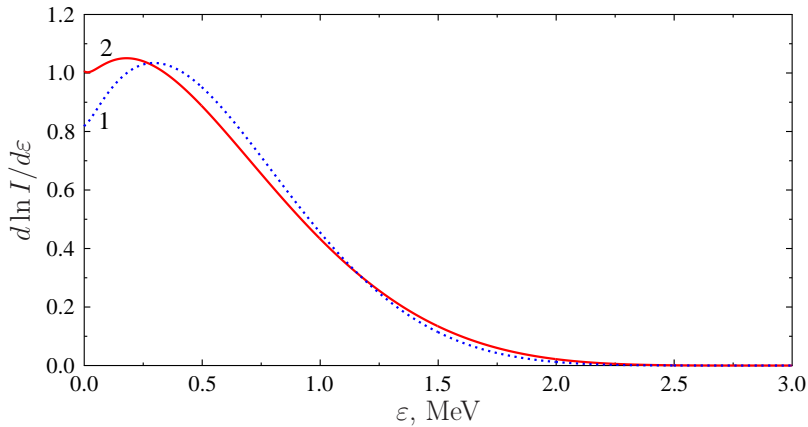}
\caption{Distribution in energy of one electron in $2\nu2\beta$-decay of
$^{82}$Se for HSD (dotted curve~1) and SSD (solid curve~2) mechanisms.}
\label{fig4}
\end{figure}

In the case of double beta decay of $^{82}$Se, the ground state of the
intermediate nucleus $^{82}$Br$_{\rm g.s.}(5^-)$ has a mass of $423$~keV
less than the mass of initial nucleus $^{82}$Se. Virtual Gamov--Teller
transition is possible through the first excited $1^+$-state of the nucleus
$^{82}$Br with $E_x=75$~keV. Accordingly,
$M(^{82}$Se$)-M(^{82}$Br$^{\ast},1_1^+)=348$~keV, and there is reason to
believe that the SSD mechanism will be realised for $2\nu2\beta$-decay of
$^{82}$Se.

The excited state of bromine-82 ($^{82}$Br$^{\ast}$, $1_1^+$) with
$E_{x}=75$~keV was found in the experiment on charge-exchange reaction
$^{82}$Se($^3$He,t)$^{82}$Br \cite{s22}, and this state is characterized by
large Gamov-Teller transition force $B(GT)=0.338$. It should be noted that the
overlying excited $1_1^+$-states of bromine-82 with $E_x<2$~MeV correspond to
the transition forces by an order of magnitude smaller. Then, basing on the
SSD hypothesis, the contribution of $1_1^+$-state of the intermediate nucleus
$^{82}$Br in the sum over $N$ in the expression (\ref{eqA2}) should only be
taken into account. Alternatively, if transition proceeds through many higher
intermediate excited states, then higher-state dominance mechanism (HSD)
governs $2\nu2\beta$-decay of $^{82}$Se. The choice of the model is the
question of physical interest, for it affects differential intensities in
two-neutrino channel, and consequently, background estimations. In favor of
SSD approach indicate the results of measurements of the intensity
distribution in the energy of one electron, which were carried out with the
NEMO-3 setup \cite{s3}. Also, investigations performed in CUPID-0 experiment
show, that SSD gives better description of the total electron energy
distribution, than HSD \cite{s2c}. NEMO-3 and future SuperNEMO detectors,
composed of a tracker and calorimeter, have capability of reconstructing of
full topology of $\beta\beta$ events. Thus a precise high-statistic study of
single-electron energy distribution, sensitive to nuclear mechanism, can be
used to distinguish between the two theoretical approaches \cite{s3}.

Nuclear matrix element $M_1^I=\langle 1_1^+\|\hat{\beta}^-\|0_i^+\rangle$
is determined from the value of the Gamov-Teller force $B(GT)=0.338$
\cite{s22} according to the relation $\left|M(GT)\right|^2=B(GT)$ that is
valid for $0^+\to 0^+$ $2\nu2\beta$-transition. Thus, $M_1^I=0.581$. It will
be possible to determine the value of matrix element
$M_1^F=\langle 0_f^+\|\hat{\beta}^-\|1_1^+\rangle$ from the study of the
recharging reaction $^{82}$Kr(d,$^2$He)$^{82}$Br, which has not yet been
completed. However, one can find $M_1^F$ with using the equations (\ref{eqA1})
and (\ref{eqA2}) on the base of the $T_{1/2}^{2\nu2\beta}$ value obtained
during the NEMO-3 experiment. In the case of SSD mechanism of
$2\nu2\beta$-decay of $^{82}$Se,
$T_{1/2}^{2\nu2\beta}=9.39\!\times\!10^{19}$~years \cite{s3}. The
corresponding value of $M_1^F$ is $0.23$, with $B(GT^+)=0.0529$.

In figure~\ref{fig4}, the distributions in the energy of one electron
corresponding to HSD and SSD mechanisms are depicted. It is certainly
interesting to compare obtained theoretical distributions with the measurement
results that will be carried out in SuperNEMO experiment.

\end{document}